\documentclass[aps,physrev,twocolumn,showpacs,amsmath,amssymb,superscriptaddress]{revtex4-2} 

\usepackage{graphicx}
\usepackage{xcolor}
\usepackage{hyperref}

\begin{document}

\title{Force and geometric signatures of the creep-to-failure transition in a granular pile}

\author{Qing Hao}
\affiliation{Department of Physics and Astronomy, Swarthmore College, Swarthmore, PA 19081, USA}

\author{Luca Montoya}
\affiliation{Department of Physics and Astronomy, Swarthmore College, Swarthmore, PA 19081, USA}

\author{Elena Lee}
\affiliation{Department of Physics and Astronomy, Swarthmore College, Swarthmore, PA 19081, USA}
\affiliation{Department of Earth and Environmental Sciences, University of Michigan, Ann Arbor, MI 48109, USA}

\author{Luke K. Davis}

\affiliation{%
School of Mathematics and Maxwell Institute for Mathematical Sciences, University of Edinburgh, Edinburgh, EH9 3FD, Scotland
}%
\affiliation{%
Higgs Centre for Theoretical Physics, University of Edinburgh, Edinburgh, EH9 3FD, Scotland
}%

\author{Cacey Stevens Bester}
\email[]{cbester1@swarthmore.edu}
\affiliation{Department of Physics and Astronomy, Swarthmore College, Swarthmore, PA 19081, USA}

\date{\today}

\begin{abstract}

Granular creep is the slow, sub-yield movement of constituents in a granular packing due to the disordered nature of its grain-scale interactions.
Despite the ubiquity of creep in disordered materials, it is still not understood how to best predict the creep-to-failure regime based on the forces and interactions among constituents.
To address this gap, we perform experiments to explore creep and failure in quasi two-dimensional piles of photoelastic disks, allowing the quantification of both grain movements and grain-scale contact force networks. 
Through controlled external disturbances, we investigate the emergence and evolution of grain rearrangements, force networks, and voids to illuminate signatures of creep and failure. 
Surprisingly, the force chain structure remains dynamic even in the absence of observable particle motion.
We find that shifts in force chains provide an indication to larger, avalanche-scale disruptions. We connect these force signatures with the geometry of the voids in the pile. 
Overall, our novel experiments and analyses deepen our mechanical and geometric understanding of the creep-to-failure transition in granular systems.

\end{abstract}

\maketitle

\textit{Introduction}. The properties of disordered materials depend on the arrangement and dynamics of their elements \cite{Behringer2018}.
Disorder produces zones within materials where irreversible deformation is shown by rearrangements of clusters of constituents. 
Large-scale and rapid flows, known as avalanches (or failure), can occur after a small disturbance \cite{amon2013}. However, the transition to failure is not abrupt. Before the point of failure, sub-yield irreversible deformation occurs in dense disordered systems at exponentially low velocities; this is known as creep \cite{komatsu2001}. Creep links seemingly disparate phenomena at different scales, from the molecular components of a glass \cite{larson, falklanger}, to the grains of a sandpile \cite{komatsu2001} and the soil along a hillside \cite{jerolmack}. Despite recent progress \cite{nicolas2018,deshpande, ferdowsi2018, amon2013}, we have an incomplete understanding of how applied stresses affect the microscopic behavior of disordered systems, which precludes our ability to predict and classify creep and failure events.
Therefore, it is of great importance to build a framework for understanding flow in disordered materials that links the individual grains, through the intermediate scale, to the system-scale. 

A comprehensive description of the flow and force transmission within a disordered system as seemingly simple as a sand pile has yet to be fully realized \cite{nicolas2018}.
The granular pile, a common configuration to explore flow of granular materials, is a classic example of how disordered materials readily lose rigidity \cite{komatsu2001,Forterre2008}.
Its constituents, large collections of dry macroscopic grains, interact via repulsive contact forces, which exhibit a complex distribution \cite{zuriguel2007}.
Static granular systems are supported heterogeneously by force chains: paths along which the strongest contact forces are carried in a network \cite{cates1998}. 
In a granular pile, force chains due to gravitational stresses sustain an ostensibly static configuration of the pile at its angle of repose, 
the critical angle at which the grains are on the verge of yield \cite{jaeger1996}. A complete description of flow transitions in granular media is one of the significant research challenges of the field of granular physics.
For example, constitutive laws of rheology--the study of flow of matter--do not fully capture sub-yield deformation in granular materials \cite{henann2014, houssais2015}.

Creep is traditionally attributed to episodic external disturbances that increase access to zones where particles are more likely to move \cite{komatsu2001, allen2018}.
However, recent work, including static granular pile simulations and experiments \cite{deshpande, ferdowsi2018} and high-precision granular rheology experiments \cite{bonn2025, yuan2024}, show the presence of creep even without external disturbances. Crucially, it is not clear which microscopic relaxation processes activate creep in granular materials, and how best to characterize them. Additionally, an open question is to understand the role of mechanical noise in causing flow below yield, as well as the analogy with noise in other disordered solids at different scales \cite{jerolmack}.

We perform tabletop granular pile experiments using photoelasticity \cite{daniels2017}, or stress-induced birefringence, to study the structural and dynamical changes due to applied stress to a granular solid near yield. A strength and novelty of this experimental approach is a high-precision microscopic description of the granular state during creep, equipped with dynamic and kinematic information. We experimentally investigate the sub-yield deformation in dry granular piles in response to external and periodic mechanical disturbances. We capture precursor rearrangement, force, and geometric events that occur before a pile has observable grain motion at the surface. Small shifts in the microstructure of granular materials have large influences in bulk behavior;
we link such microscale and mesoscale changes in particle positions, force fluctuations, and voids, to the macroscale continuum flow.

% experimental setup figure
\begin{figure}[h]
\centering
\includegraphics[width = 0.9
\linewidth]{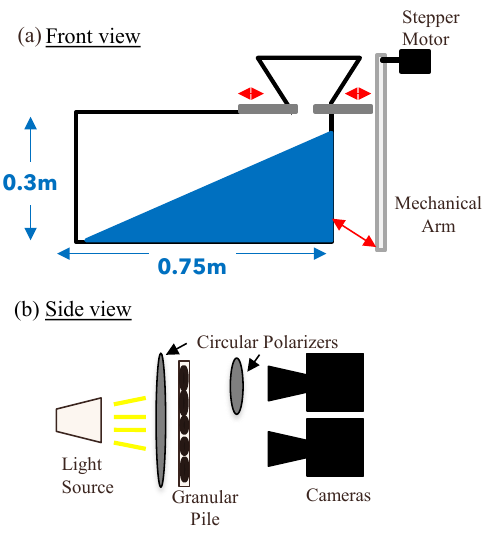}
\caption{\textbf{Experimental apparatus:}
(a) Granular piles are constructed using the localized source procedure, as grains flow through an attached hopper at 40 cm above the surface to form a heap geometry. External disturbances to the pile result from a mechanical arm periodically tapping the acrylic container. 
(b) Side view shows grains illuminated by circularly polarized light and imaged simultaneously with two cameras to resolve force chains (using a second circular polarizer) and grain positions of a photoelastic granular system.  
}
\label{fig:apparatus}
\end{figure}

\textit{Experimental setup}. 
The experimental apparatus (Fig. \ref{fig:apparatus}) is composed of a quasi-two-dimensional transparent acrylic chamber with dimensions of 75 cm wide and 30 cm high.
For granular media, we use a bidisperse arrangement of about 1500 disks of 6 mm and 9 mm diameter in a 1:1 number ratio.
Grains were cast from Clearflex 50 urethane using a previously outlined procedure \cite{abedzadeh2019}.
The grains are placed in a nearly vertical plane, and we dust the grains with talc powder to improve lubrication and reduce friction with side walls.

Granular piles are formed using the localized source procedure \cite{zuriguel2007}, in which grains flow through a double hopper above the base of the apparatus.
The preparation of the pile plays a critical role in its resulting stress profile at the base.
Once the grains are poured into the pile, we wait approximately thirty minutes before beginning the experimental observation.
We investigate the pile formation process to observe the evolution of the contact force network. The pile settles to a specific internal structure which, upon inspection of the particle displacements and force chains, does not have observable change up to the start of the experimental run (see Supplemental Material \cite{suppl_material} Fig. 1 for pile relaxation measurements).

The key to these experiments is that the bidisperse grains are made from a material that exhibits stress-induced birefringence, or photoelasticity \cite{hecht}.
This allows for both acquisition of grain positions and imaging of the contact force network, providing a complete description of the granular state during sub-yield deformation.
When a photoelastic material is placed between two perpendicularly oriented circular polarizers and subjected to stress, regions of the material alter the polarization of light. 
The result is a visual pattern of bright grains within the system, which allows us to observe the local stress in each region of the pile \cite{daniels2017, abedzadeh2019}.
This method thus provides a direct visualization of contact forces and their evolution under flow among granular materials.

\begin{figure}[h]
\centering
\includegraphics[width=0.85\linewidth]{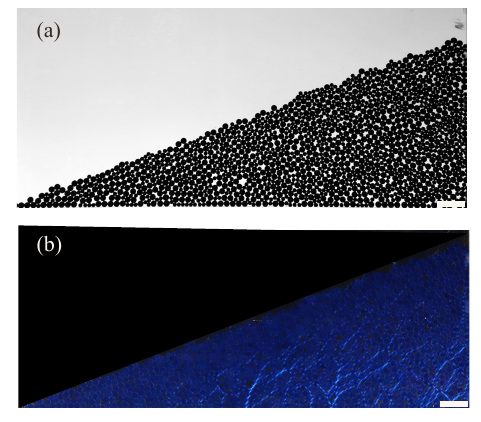}
\caption{\textbf{Photoelastic granular pile:}
Two cameras simultaneously capture images
of a photoelastic granular system in a heap geometry as it is periodically disturbed via tapping. 
(a) Grains are imaged with lighting without crossed polarizers.  
(b) Force chains, seen as lightning-like patterns in the image, are revealed when viewed between crossed circular polarizers. The scale bars are 40 mm.
}
\label{fig:expt}
\end{figure}

To capture creep and failure events, we use two high-resolution cameras that simultaneously record images of the full granular system approximately every three seconds.
This dual-camera setup allows us to  gather information about the grain positions, as in Fig. \ref{fig:expt}(a), and the underlying force chains, as shown in Fig. \ref{fig:expt}(b). 
To determine grain positions, we apply a circular Hough transform to the experimental images via the MATLAB function \textit{imfindcircles}, achieving a positional accuracy within 2 pixels.
Images are mapped onto each other during post-processing, enabling a straightforward coupling of the evolution of particle arrangements and force chain structures. We employ the photoelastic technique to make semiquantitative measurements of contact forces as recently used to study granular flows on the free surface \cite{thomas2019}. 
Force chains refer to the chain-like distributions of grains that bear higher than average loads through the pile and illustrate heterogeneous force transmission as a fundamental feature of granular media \cite{cates1998}.
They are shown as a result of the weight of the grains above them and depict the grains that maintain the strongest load.

It is important to determine how global external disturbances influence creep.
Examples of natural disturbances are seismic events that cause slow changes in landscapes and landslide processes \cite{jerolmack}.
Here, we introduce periodic disturbances every ten seconds via tapping to piles that are initially in an unperturbed solid-like state.
We employ a mechanical arm, driven by a stepper motor, to tap the acrylic chamber at intervals.
Taps are applied to the lower row of grains of the pile.
This provides an inflow of kinetic energy to the system.
For the applied force, tapping accelerates the rate of creep in the system \cite{deshpande}. 
As the granular pile is tapped, we characterize the packing and force chain structure using image analysis.

\begin{figure}[ht!]
\includegraphics[width = 0.99\linewidth]{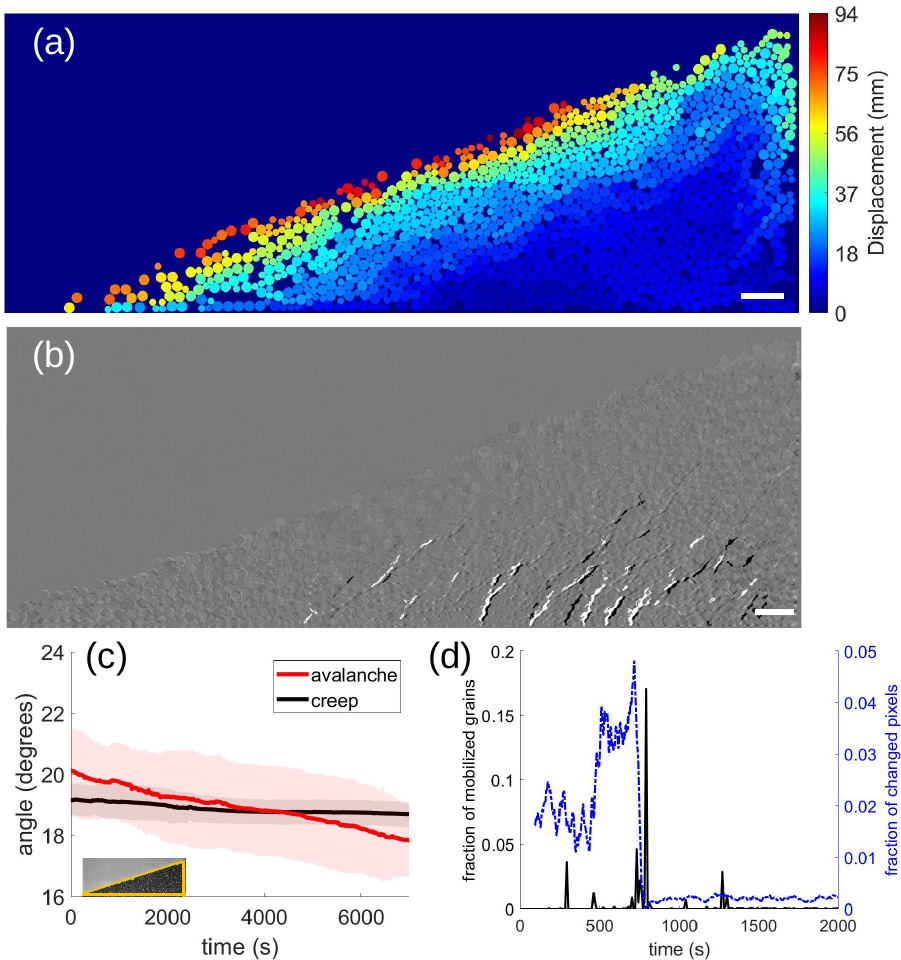}
\caption{ \textbf{Structure and dynamics during creep}: 
(a) Particle displacements are labelled by color onto initial grain positions to illustrate grain motion 
throughout the pile over two hours of observation.
(b) Difference of photoelastic images. An image at the start of an avalanche is subtracted from images taken before an avalanche. 
Changes in force chains are visible within the bulk of the pile and marked by black or white pixels. 
(c) Pile surface angle versus time for creeping and avalanching piles found from the perimeter of the pile (see inset). The solid lines are averages of runs and standard deviations are represented by the shaded regions.
(d) Fraction of black and white pixels as determined from difference images plotted with number of grains that move above a distance threshold of one particle radius, showing how stress redistribution leads to creep and failure.
Piles are tapped every 10 seconds. Scale bars are 40mm.
}
\label{fig:difference}
\end{figure}

\textit{Results}. From the experiments, we clearly observe the distribution of grains 
relative to their displacement from their initial positions in the pile (see Fig. \ref{fig:difference}(a)). 
Over the course of an experimental run, there is grain motion at the surface (average speed over full run $\approx 1.4\times 10^{-5}$~m/s) and deep within the pile (average speed $\approx 2\times 10^{-6}$~m/s). We observe multi-grain flow events after shifts of a single grain (see Supplemental Material \cite{suppl_material} for video). After suitable thresholding of grain displacements according to time and distance measures, we label surface flow events as avalanches, in which there are about 10\% of grains that move as accompanied with dynamic and geometric changes. Accordingly, we highlight two qualitative flow phases: creep and failure. 
In this context, we define failure as the onset of surface flow, or an avalanche, in an initially stationary pile. 
To capture the slight changes that occur in the force network, we analyze the difference between photoelastic images (see Fig. \ref{fig:difference}(b)) \cite{utter2013}. Qualitatively, the black and white pixels in these images capture the locations of contact force migration. From these, we observe changes in contact arrangements as creep destabilizes the force network. 

To quantify macroscale flow behavior, we track the temporal evolution of the pile angle, comparing the slopes of creeping piles to those that experience failure (Fig. \ref{fig:difference}(c)).
The pile naturally forms at an angle for which the pile is the most fragile and is thereby susceptible to creep. 
Through a separate experiment, we determine the angle to be $20-24^\circ$. This is achieved by pouring grains to form a conical pile within an acrylic box and recording the structural state immediately preceding the first surface avalanche.
As expected, the surface angle of creeping piles decreases only marginally over time. In contrast, piles that undergo avalanching begin with a higher average initial angle—closer to the angle of repose—and subsequently collapse to a final angle lower than that of typical creeping piles. Notably, this sub-creep final angle implies that macroscopic failure events allow the pile to reorganize into a distinctly more stable configuration, \emph{i.e.,} negligible fraction of mobilized grains (as in SM Fig. 1).

We quantify the coupling between force network dynamics and localized grain rearrangements by tracking their time-series evolution during creep phases that ultimately transition into avalanches (Fig. \ref{fig:difference}(d)). 
In this analysis, grains are defined as mobilized if they displace by at least one small-grain radius between successive frames. 
Our results demonstrate that the fraction of altered pixels within the force network varies tenfold before the mobilized grain fraction reaches a peak of 0.17. Crucially, force chain reconfiguration occurs even without detectable grain kinematics, establishing stress redistribution as an essential signature of creep and failure. 
This simultaneous measurement of internal forces and microstructural flow provides key physical insights into macroscopic granular failure.

\begin{figure}[t!]
\centering
\includegraphics[width=\linewidth]{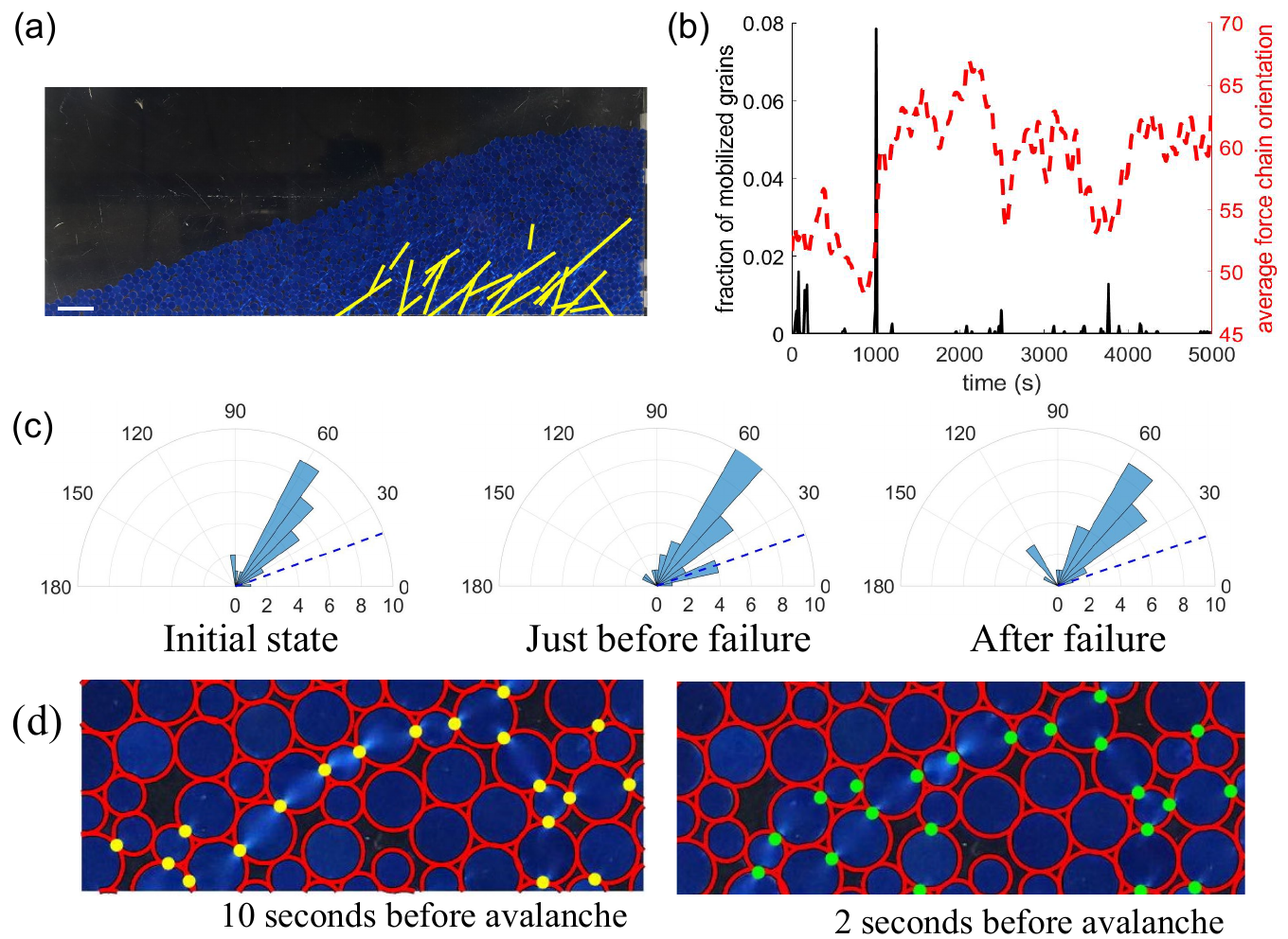}
\caption{ \textbf{Force chains give way to creep}:
(a) Image taken under polarized light with force chains above a length and brightness threshold labeled using a line segment (see main text). The scale bar is 40 mm. 
(b) The average force chain orientation is plotted with the number of mobilized grains as a function of time.
(c) The distribution of force chain orientations are shown at the initial state of an experimental run, as the pile creeps, and just after failure. 
The blue dashed line is the angle of the pile surface at that instant.
(d) Zoomed-in view of the contact network at different times during pile creep. 
}
\label{fig:forces}
\end{figure}

Further image analysis enables the identification and characterization of force chains within the granular pile (Fig. \ref{fig:forces}(a)). This image analysis relies on sequential intensity and length thresholding.  Following contrast enhancement, the polarized light images are binarized using a global intensity threshold ($>30\%$ of background value) indicating higher-than-average load transmission. The length-thresholding consists of a connected-component analysis of grain contact points applied to the binary images where we retain only structures longer than five small-grain radii ($> 30$~mm). This length scale is chosen to ensure each identified chain encompasses at least two grain contacts. Having specified the thresholding procedure, force chains emerge as line segments of pixels that connect contact points. This typically results in approximately linear force chains (Fig. \ref{fig:forces}). 

Significant shifts in force chain orientations coincide directly with changes in the fraction of mobilized grains (Figs. \ref{fig:forces}(b) and \ref{fig:difference}(d)). 
This shift in average orientation underscores the stress redistribution that acts as a precursor to macroscopic grain flow. As shown in Fig. \ref{fig:forces}(c), primary force chains are oriented predominantly at 50°–65° relative to the horizontal, accompanied by a secondary network of perpendicularly oriented chains.  This analysis demonstrates that force structure reorganization during creep precipitates larger-scale flow events. Immediately prior to failure, the distribution of force chain orientations broadens, exhibiting a noticeable increase in alignments parallel to the surface of the pile that facilitate the impending avalanche. Following failure, these surface-parallel orientations are suppressed, resulting in an overall increase in the average force chain orientation.

We investigate how grain-scale contacts relate to the dynamics of the force chain network (Fig. \ref{fig:forces}d) \cite{majmudar2005}. 
Because this network dictates the mechanical stability of the transitioning granular system, tracking its rearrangement during creep reveals the microscale measurements of stress redistribution before avalanches. 
We define the contact by identifying interparticle connections that satisfy the photoelastic intensity threshold. 
To determine contacts, we locate grain centers and boundaries using unpolarized light images then map these positions onto corresponding photoelastic images to evaluate the force signals between neighboring grains. 
As the pile approaches failure, we find that stress redistribution occurs without grain displacement within experimental resolution. 
For example, Fig. \ref{fig:forces}(d) illustrates two time steps where particle positions remain static within experimental resolution, yet the underlying force network clearly evolves. This evolution manifests as the continuous formation and loss of contacts, exposed by local changes in photoelastic intensity. 
Furthermore, even a single grain rearrangement deep within the bulk can significantly alter the local force distribution, weakening the regional contact network and potentially nucleating larger flow events.

{The arrangements of the constituents in the granular pile are expected to influence creep and failure, but it is not clear how structure relates to deformation and the force network \cite{Cubuk2017}}. Encouraged by previously successful applications of the concepts of void space in granular media \cite{Knight1995,Boutreux1997,Nowak1998,Lemaitre2002,Behringer2018, ramos2009}, we explore the amount and distribution of unoccupied volume (voids) to improve the understanding of the creep-to-failure transition in granular systems.

We define the total unoccupied volume, $V_u$, in the pile as:
\begin{equation}
    V_u := \sum_{m=1}^{N_\Delta(t)} \Delta_m,
    \label{eq:voids}
\end{equation}
where $N_\Delta(t)$ is the instantaneous number of voids with the $m^\text{th}$ void having a two-dimensional volume of $\Delta_m$. 
The advantage of determining $V_u$ via microscopic voids is to access particle-level structural information, particularly after rearrangement events in which some of the voids are expected to change size. 
We hypothesize that these rearrangements result in changes in the number of voids $N_\Delta$, the overall distribution of $\Delta$, and the total unoccupied volume $V_u$. 

\begin{figure}[h]
    \centering
    \includegraphics[width=0.75\linewidth]{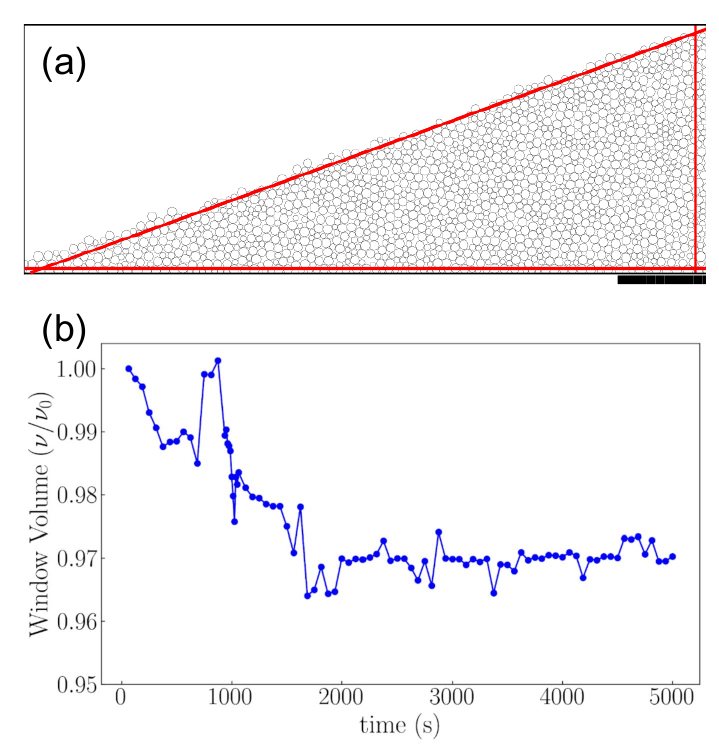}
    \caption{\textbf{Set up for void analysis:} (a) Example of extracted grain positions and sizes with a triangle (red lines) fitted to constrain the void analysis spatial region.  (b) Values of the calculation triangle volume as a function of time, divided by its initial value, for a pile dataset containing an avalanche event around 1000~s.}
    \label{fig:triangle}
\end{figure}

To determine the unoccupied volume, we develop a workflow which involves first defining a triangular (focus) region, voxelizing the space in this region, and excluding voxels that overlap with the grains.
Specifically, we used the following criterion to exclude voxels:
\begin{equation}
    \begin{aligned}
        \text{Exclude if} \qquad |\mathbf{R}_i - \mathbf{r}_m | \leq l + \sigma_i,
    \end{aligned}
\end{equation}
where $\mathbf{R}_i$ is the position of grain $i$, $\mathbf{r}_m$ is the position of voxel $m$, $l$ is the voxel side-length, and $\sigma_i$ is the diameter of grain $i$. 
We choose $l = 0.375$~mm to resolve individual void volumes $\Delta_m$.
This length was deemed small enough to capture the shape of the cavities and large enough to be computationally efficient.
Once the voxels have been excluded, a standard hierarchical agglomeration clustering algorithm is then used to group together adjacent voxels.

The triangular region consists of a diagonal line $y=mx+c$, with $m = y_1-y_0/ x_1 - x_0$ determined from positions of the leftmost-lowest grain $(x_0,y_0)$ and the rightmost-highest grain $(x_1,y_1)$, a vertical line at $x_\text{high}$, and a horizontal line at $y_\text{low}$ (see fig. \ref{fig:triangle}(a)). 
The parameters $c, y_\text{low}$, and $x_\text{high}$ are adjusted incrementally until $95\%$ of the grain positions are contained.
The total volume of the calculation triangle is given as:
\begin{equation}
    \nu = V_\text{discs} + V_u,
\end{equation}
where $V_\text{discs}$ is the volume of the discs. 
Thus, knowing $\nu$ and $V_\text{discs}$, one is able to obtain the total void space $V_u$ for comparison with eq. (1).
As the unoccupied volume decreases we see the volume of the calculation triangle decrease similarly (see Fig. \ref{fig:triangle}(b)).
We note that the bulk of the pile can only get denser (lower unoccupied volume) if and only if particles within the bulk rearrange so as to expel unoccupied volume from this region onto the free surface of the pile.
The peak in $\nu$ near $t = 1000$ s serves as a characteristic signature of the avalanche, reflecting system dilation during the creep-to-failure transition.

% voids figure figure
\begin{figure}[t!]
\centering
\includegraphics[width=\linewidth]{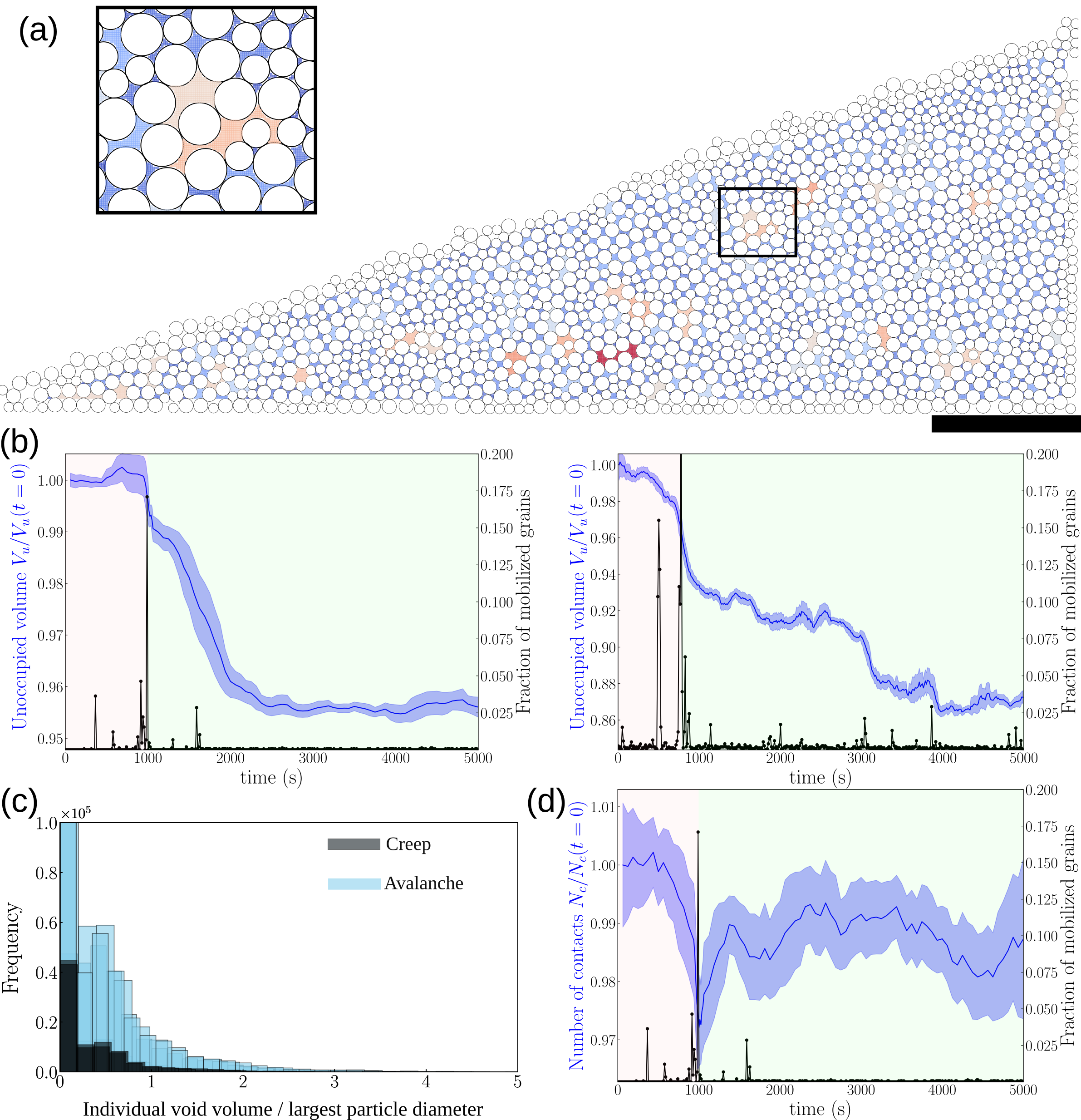}
\caption{ \textbf{Geometric signatures of rearrangements in the unoccupied volume:} (a) Snapshot of the voids in a pile, calculated using an efficient voxel-based exclusion algorithm which uses positions and diameters of the grains.
(b) Quantitative comparison from two trials between the unoccupied volume $V_u$ (divided by its initial value) and the fraction of mobilized grains as a function of time. The solid line represents a smoothed-out average, and the shaded regions provide mean prediction bands (95\%).
(c) Histograms of individual void sizes taken from independent creep (grey) and avalanche (blue) data during all time steps. (d) Averaged grain contact numbers as a function of time for the same avalanche dataset shown in (left, b). Scale bar in (a) is 90~mm.}
\label{fig:voids}
\end{figure}

Figure \ref{fig:voids}(a) highlights the two-dimensional voids in a granular pile.
Surprisingly, we find that the unoccupied volume contains signatures of avalanching events (see Figs. \ref{fig:voids}(b) and Supplemental Material \cite{suppl_material} Fig. 4(a)). 
We observe that grain displacement events coincide with shifts in the unoccupied volume. 
In the avalanche data we observe a tendency towards lower unoccupied volume, which can be explained microscopically: during rearrangements, grains tend to move to fill larger voids which ultimately reduces the total unoccupied volume. 
The gradual decrease of the unoccupied volume implies that the pile is becoming more stable, which is consistent with Fig. \ref{fig:difference}(c).
To further test this observation, we analyze the void space as a function of time in a pile formation process where large grain rearrangements occur at the beginning (due to the deposition of grains) and negligible grain motion is observed at later times (absence of external disturbances). 
The unoccupied volume decreases immediately following grain deposition, stabilizing as the granular pile relaxes.

Importantly, we find that the unoccupied volume could distinguish between creep and failure events, where the overall distribution of the void sizes is smaller in the creep data (see Fig. \ref{fig:voids}(c)). 
To better connect void volumes and forces we also computed the average contact number (see Fig. \ref{fig:voids}(d)). 
 For each particle $i$, a contact is defined through the following relation being true
\begin{equation}
    |\mathbf{R}_i - \mathbf{R}_j| \leq \frac{D_i + D_j}{2}, 
\end{equation}
where $D_i$ is the diameter of particle $i$. To avoid comparing every particle with every other particle, we implement a cell-based method with a cell length of $D_\text{max}/2$ where $D_\text{max}$ is the largest particle diameter. Thus, only particles in the neighboring cells need to be checked.
A reduction in interparticle contacts coincides with a slight increase in unoccupied volume immediately prior to the avalanche (Fig. \ref{fig:voids}(d)). 
This localized dilation reduces constraints, granting particles greater mobility and destabilizing the pile. 
Consequently, this coupled behavior serves as a distinct microscale precursor to macroscopic failure.

\begin{figure}[h]
\centering
\includegraphics[width=0.8\linewidth]{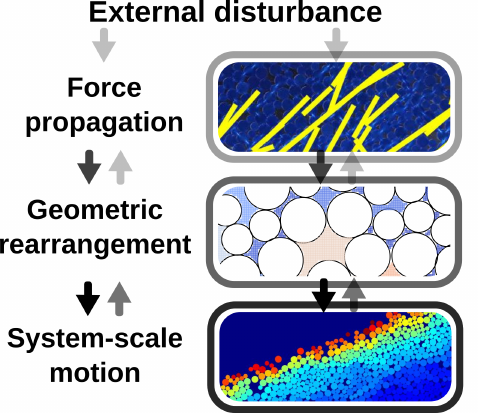}
\caption{ \textbf{Multi-scale understanding of creep-to-failure in granular piles:} Pile behavior can be described through signatures between force propagation, geometric rearrangements, and system-scale motion.}
\label{fig:takehome}
\end{figure}

\textit{Conclusion}. 
In this study, we experimentally investigate the conditions under which a granular pile creeps and ultimately fails by characterizing the evolution of grain displacements, force chain evolution, and void distributions. 
Our data reveal correlations between grain displacements and dynamic structural reorganization (Figs. \ref{fig:difference}, \ref{fig:forces}, and Fig. \ref{fig:voids}), allowing us to map dynamic and geometric signatures to macroscopic yielding (Fig. \ref{fig:takehome}). 
Notably, we demonstrate that dynamic structural reorganization occurs deep within the bulk of the pile, rather than being confined solely to the free surface. 
Furthermore, our analysis of the creep phase revealed that the force contact network remains dynamic even when grain motion is below the resolution of measurement.

By tracking localized changes in forces and geometry at the microscale and mesoscale immediately preceding an avalanche, we aim to elucidate the physical origins of macroscopic failure. Specifically, we propose that the configuration of microscopic voids and the structure of mesoscale force chains actively drive these transitions to flow.
Ultimately, this work reveals a profound coupling between mechanics and geometry in yielding granular media. 
By capturing these signatures, we advance towards a framework that explains how the interplay between mechanics and geometry during creep drives the onset of macroscopic failure. A key remaining challenge lies in arriving at mathematical relationships that link motion, force networks, and geometry in a granular pile.
\nocite{vanel1999}

\textit{Data availability}. The data, besides those in the article itself, and software used in this study are available upon reasonable request.

\textit{Acknowledgements}. The work was supported by the Research Corporation for Science Advancement, American Physical Society Innovation Fund, and Swarthmore College Provost's Office.
C.S.B. is particularly thankful to Douglas Jerolmack for valuable discussions.
We thank Paul Jacobs and Steven Palmer for assistance with apparatus development, and Ben McMillian, Greg Voth, and Nathalie Vriend for helpful discussions.
The Squishlab at Haverford College, led by Ted Brzinski, provided grains. L.K.D. acknowledges funding from the Flora Philip Fellowship at the University of Edinburgh.
% \end{acknowledgments}

\bibliography{references.bib}

@article{Behringer2018,
  title = {The physics of jamming for granular materials: a review},
  volume = {82},
  ISSN = {1361-6633},
  url = {http://dx.doi.org/10.1088/1361-6633/aadc3c},
  DOI = {10.1088/1361-6633/aadc3c},
  number = {1},
  journal = {Reports on Progress in Physics},
  publisher = {IOP Publishing},
  author = {Behringer,  Robert P and Chakraborty,  Bulbul},
  year = {2018},
  month = nov,
  pages = {012601}
}

@ARTICLE{lemaitre2002,
  title     = "Rearrangements and dilatancy for sheared dense materials",
  author    = "Lema{\^\i}tre, Ana{\"e}l",
  journal   = "Phys. Rev. Lett.",
  publisher = "American Physical Society (APS)",
  volume    =  89,
  number    =  19,
  pages     = "195503",
  month     =  nov,
  year      =  2002,
  copyright = "http://link.aps.org/licenses/aps-default-license",
  language  = "en"
}

@article{Boutreux1997,
  title = {Compaction of granular mixtures: a free volume model},
  volume = {244},
  ISSN = {0378-4371},
  url = {http://dx.doi.org/10.1016/S0378-4371(97)00236-7},
  DOI = {10.1016/s0378-4371(97)00236-7},
  number = {1–4},
  journal = {Physica A: Statistical Mechanics and its Applications},
  publisher = {Elsevier BV},
  author = {Boutreux,  T. and de Geennes,  P.G.},
  year = {1997},
  month = oct,
  pages = {59–67}
}

@article{Nowak1998,
  title = {Density fluctuations in vibrated granular materials},
  volume = {57},
  ISSN = {1095-3787},
  url = {http://dx.doi.org/10.1103/PhysRevE.57.1971},
  DOI = {10.1103/physreve.57.1971},
  number = {2},
  journal = {Physical Review E},
  publisher = {American Physical Society (APS)},
  author = {Nowak,  Edmund and Knight,  James and Ben-Naim,  Eli and Jaeger,  Heinrich and Nagel,  Sidney},
  year = {1998},
  month = feb,
  pages = {1971–1982}
}

@article{Cubuk2017,
  title = {Structure-property relationships from universal signatures of plasticity in disordered solids},
  volume = {358},
  ISSN = {1095-9203},
  url = {http://dx.doi.org/10.1126/science.aai8830},
  DOI = {10.1126/science.aai8830},
  number = {6366},
  journal = {Science},
  publisher = {American Association for the Advancement of Science (AAAS)},
  author = {Cubuk,  E. D. and Ivancic,  R. J. S. and Schoenholz,  S. S. and Strickland,  D. J. and Basu,  A. and Davidson,  Z. S. and Fontaine,  J. and Hor,  J. L. and Huang,  Y.-R. and Jiang,  Y. and Keim,  N. C. and Koshigan,  K. D. and Lefever,  J. A. and Liu,  T. and Ma,  X.-G. and Magagnosc,  D. J. and Morrow,  E. and Ortiz,  C. P. and Rieser,  J. M. and Shavit,  A. and Still,  T. and Xu,  Y. and Zhang,  Y. and Nordstrom,  K. N. and Arratia,  P. E. and Carpick,  R. W. and Durian,  D. J. and Fakhraai,  Z. and Jerolmack,  D. J. and Lee,  Daeyeon and Li,  Ju and Riggleman,  R. and Turner,  K. T. and Yodh,  A. G. and Gianola,  D. S. and Liu,  Andrea J.},
  year = {2017},
  month = nov,
  pages = {1033–1037}
}

@article{Knight1995,
  title = {Density relaxation in a vibrated granular material},
  volume = {51},
  ISSN = {1095-3787},
  url = {http://dx.doi.org/10.1103/PhysRevE.51.3957},
  DOI = {10.1103/physreve.51.3957},
  number = {5},
  journal = {Physical Review E},
  publisher = {American Physical Society (APS)},
  author = {Knight,  James B. and Fandrich,  Christopher G. and Lau,  Chun Ning and Jaeger,  Heinrich M. and Nagel,  Sidney R.},
  year = {1995},
  month = may,
  pages = {3957–3963}
}

@ARTICLE{yuan2024,
  title     = "From creep to flow: Granular materials under cyclic shear",
  author    = "Yuan, Ye and Zeng, Zhikun and Xing, Yi and Yuan, Houfei and
               Zhang, Shuyang and Kob, Walter and Wang, Yujie",
  journal   = "Nat. Commun.",
  publisher = "Springer Science and Business Media LLC",
  volume    =  15,
  number    =  1,
  pages     = "3866",
  month     =  may,
  year      =  2024,
  copyright = "https://creativecommons.org/licenses/by/4.0",
  language  = "en"
}

@ARTICLE{thomas2019,
  title     = "Photoelastic study of dense granular free-surface flows",
  author    = "Thomas, A L and Vriend, N M",
  journal   = "Phys. Rev. E.",
  publisher = "American Physical Society (APS)",
  volume    =  100,
  number    = "1-1",
  pages     = "012902",
  month     =  jul,
  year      =  2019,
  copyright = "https://link.aps.org/licenses/aps-default-license",
  language  = "en"
}

@ARTICLE{daniels2017,
  title     = "Photoelastic force measurements in granular materials",
  author    = "Daniels, Karen E and Kollmer, Jonathan E and Puckett, James G",
  journal   = "Rev. Sci. Instrum.",
  publisher = "AIP Publishing",
  volume    =  88,
  number    =  5,
  pages     = "051808",
  month     =  may,
  year      =  2017,
  language  = "en"
}

@ARTICLE{abedzadeh2019,
  title     = "Enlightening force chains: a review of photoelasticimetry in
               granular matter",
  author    = "Abed Zadeh, Aghil and Bar{\'e}s, Jonathan and Brzinski, Theodore
               A and Daniels, Karen E and Dijksman, Joshua and Docquier,
               Nicolas and Everitt, Henry O and Kollmer, Jonathan E and
               Lantsoght, Olivier and Wang, Dong and Workamp, Marcel and Zhao,
               Yiqiu and Zheng, Hu",
  journal   = "Granul. Matter",
  publisher = "Springer Science and Business Media LLC",
  volume    =  21,
  number    =  4,
  month     =  nov,
  year      =  2019,
  language  = "en"
}

@book{larson,
    author = {R. G. Larson},
    title = {The Structure and Rheology of Complex Fluids},
    publisher = {Oxford University Press},
    year = {1999},
}

@article{Forterre2008,
  title = {Flows of Dense Granular Media},
  volume = {40},
  ISSN = {1545-4479},
  url = {http://dx.doi.org/10.1146/annurev.fluid.40.111406.102142},
  DOI = {10.1146/annurev.fluid.40.111406.102142},
  number = {1},
  journal = {Annual Review of Fluid Mechanics},
  publisher = {Annual Reviews},
  author = {Forterre,  Yoël and Pouliquen,  Olivier},
  year = {2008},
  month = jan,
  pages = {1–24}
}

@ARTICLE{falklanger,
  title     = "Dynamics of viscoplastic deformation in amorphous solids",
  author    = "Falk, M L and Langer, J S",
  journal   = "Phys. Rev. E Stat. Phys. Plasmas Fluids Relat. Interdiscip.
               Topics",
  publisher = "American Physical Society (APS)",
  volume    =  57,
  number    =  6,
  pages     = "7192--7205",
  month     =  jun,
  year      =  1998,
  copyright = "http://link.aps.org/licenses/aps-default-license"
}

@article{jerolmack,
    author = {D. J. Jerolmack and K. E. Daniels},
   doi = {},
   issn = {},
   journal = {Nature Reviews Physics},
   title = {Viewing Earth’s surface as a soft-matter landscape},
   volume = {1:12},
   year = {2019},
}

@article{Komatsu2001,
  title = {Creep Motion in a Granular Pile Exhibiting Steady Surface Flow},
  volume = {86},
  ISSN = {1079-7114},
  url = {http://dx.doi.org/10.1103/PhysRevLett.86.1757},
  DOI = {10.1103/physrevlett.86.1757},
  number = {9},
  journal = {Physical Review Letters},
  publisher = {American Physical Society (APS)},
  author = {Komatsu,  Teruhisa S. and Inagaki,  Shio and Nakagawa,  Naoko and Nasuno,  Satoru},
  year = {2001},
  month = feb,
  pages = {1757–1760}
}

@article{deshpande,
    author = {N. Deshpande and D. J. Furbish and P. E. Arratia and D. J. Jerolmack},
    title = {The perpetual fragility of creeping hillslopes},
    journal = {Nature Communications},
   volume = {12:1},
    year = {2021}
}

@article{bonn2025,
    author = {Kasra Farain and Daniel Bonn},
    title = {Anomalous Creep as a Precursor to Failure in Granular Materials},
    journal = {arXiv preprint arXiv:2502.02288},
   volume = {},
    year = {2025}
}

@article{zuriguel2007,
    author = {I. Zuriguel and T. Mullin and J.M. Rotter},
    title = {Effect of particle shape on the stress dip under a sandpile},
    journal = {Physical Review Letters},
   volume = {98, 028001},
    year = {2007}
}

@article{utter2013,
    author = {D.L. Amon and T. Niculescu and B.C. Utter},
    title = {Granular avalanches in a two-dimensional rotating drum with imposed vertical vibration},
    journal = {Physical Review E},
   volume = {88(1), 012203},
    year = {2013}
}

@article{henann2014,
    author = {D. L. Henann and K. Kamrin},
    title = {Continuum modeling of secondary rheology in dense granular materials},
    journal = {Physical Review Letters},
   volume = {113:17},
    year = {2014}
}

@article{houssais2015,
    author = {M. Houssais and C. P. Ortiz and D. J. Durian and D. J. Jerolmack},
    title = {Onset of sediment transport is a continuous transition driven by fluid shear and granular creep},
    journal = {Nature Communications},
   volume = {6:1},
    year = {2015}
}

@article{cates1998,
    author = {M. E. Cates and J. P. Wittmer and J. P. Bouchard and P. Claudin},
    title = {Jamming, force chains, and fragile matter},
    journal = {Physical Review Letters},
   volume = {81:9},
    year = {1998}
}

@article{jaeger1996,
    author = {H. M. Jaeger and S. R. Nagel and R. P. Behringer},
    title = {Granular Solids, Liquids, and Gases},
    journal = {Reviews of Modern Physics},
   volume = {68},
issn = {4},
    year = {1996}
}

@ARTICLE{allen2018,
  title     = "Granular bed consolidation, creep, and armoring under
               subcritical fluid flow",
  author    = "Allen, Benjamin and Kudrolli, Arshad",
  journal   = "Phys. Rev. Fluids",
  publisher = "American Physical Society (APS)",
  volume    =  3,
  number    =  7,
  month     =  jul,
  year      =  2018,
  copyright = "https://link.aps.org/licenses/aps-default-license",
  language  = "en"
}

@article{amon2013,
    author = {A. Amon and R. Bertoni and J. Crassous},
    title = {Experimental investigation of plastic deformations before a granular avalanche},
    journal = {Physical Review E},
   volume = {87},
    issn = {1},
    year = {2013}
}

@article{ferdowsi2018,
    author = {B. Ferdowsi and C. P. Ortiz and D. J. Jerolmack},
    title = {Glassy dynamics of landscape evolution},
    journal = {Proceedings of the National Academy of Sciences},
   volume = {115},
    issn = {19},
    year = {2018}
}

@book{hecht,
  author = {Eugene Hecht},
  year = {2002},
  title = {Optics},
  publisher = {Addison-Wesley Professional}
}

@ARTICLE{majmudar2005,
  title     = "Contact force measurements and stress-induced anisotropy in
               granular materials",
  author    = "Majmudar, T S and Behringer, R P",
  journal   = "Nature",
  publisher = "Springer Science and Business Media LLC",
  volume    =  435,
  number    =  7045,
  pages     = "1079--1082",
  month     =  jun,
  year      =  2005,
  language  = "en"
}

@ARTICLE{vanel1999,
  title     = "Memories in sand: Experimental tests of construction history on
               stress distributions under sandpiles",
  author    = "Vanel, Loic and Howell, Daniel and Clark, D and Behringer, R P
               and Cl{\'e}ment, Eric",
  journal   = "Phys. Rev. E Stat. Phys. Plasmas Fluids Relat. Interdiscip.
               Topics",
  publisher = "American Physical Society (APS)",
  volume    =  60,
  number    =  5,
  pages     = "R5040--R5043",
  month     =  nov,
  year      =  1999,
  copyright = "http://link.aps.org/licenses/aps-default-license"
}

@ARTICLE{nicolas2018,
  title     = "Deformation and flow of amorphous solids: Insights from
               elastoplastic models",
  author    = "Nicolas, Alexandre and Ferrero, Ezequiel E and Martens, Kirsten
               and Barrat, Jean-Louis",
  journal   = "Rev. Mod. Phys.",
  publisher = "American Physical Society (APS)",
  volume    =  90,
  number    =  4,
  month     =  dec,
  year      =  2018,
  copyright = "https://link.aps.org/licenses/aps-default-license",
  language  = "en"
}

@article{ramos2009,
  title={Avalanche prediction in a self-organized pile of beads},
  author={Ramos, Osvanny and Altshuler, E and M{\aa}l{\o}y, KJ},
  journal={Physical Review Letters},
  volume={102},
  number={7},
  pages={078701},
  year={2009},
  publisher={APS},
  doi={10.1103/PhysRevLett.102.078701}
}

@misc{suppl_material,
  note = {See Supplemental Material at https://link.aps.org/supplemental/10.1103/vjg6-xb7p for further detail on pile formation, disturbance profiles, and additional data.}
}

\end{document}

\title{Supplemental material: Force and geometric signatures of the creep-to-failure transition in a granular pile}

\author{Qing Hao}
\affiliation{Department of Physics and Astronomy, Swarthmore College, Swarthmore, PA 19081, USA}

\author{Luca Montoya}
\affiliation{Department of Physics and Astronomy, Swarthmore College, Swarthmore, PA 19081, USA}

\author{Elena Lee}
\affiliation{Department of Physics and Astronomy, Swarthmore College, Swarthmore, PA 19081, USA}
\affiliation{Department of Earth and Environmental Science, University of Michigan, Ann Arbor, MI 48109, USA}

\author{Luke K. Davis}

\affiliation{%
School of Mathematics and Maxwell Institute for Mathematical Sciences, University of Edinburgh, Edinburgh, EH9 3FD, Scotland
}%
\affiliation{%
Higgs Centre for Theoretical Physics, University of Edinburgh, Edinburgh, EH9 3FD, Scotland
}%

\author{Cacey Stevens Bester}
\email[]{cbester1@swarthmore.edu}
\affiliation{Department of Physics and Astronomy, Swarthmore College, Swarthmore, PA 19081, USA}

\date{\today}

\maketitle

\newpage
\section{Pile formation}

The grains are poured into a quasi-2D rectangular cell from a hopper.
We form the pile using a localized source procedure in which the hopper outlet is much smaller than the final pile diameter \cite{vanel1999}.
We record the pile relaxation using the dual-camera setup, acquiring images every three seconds.
The data shown in supplemental fig. \ref{fig:pile_formation} are representative of each experimental run.
We track the displacements of mobilized grains by connecting grain positions across sequential images.
We treat a grain as mobilized if it moves at least one small grain radius (22 pixels).
After about 600 seconds, the pile has settled to a specific internal structure which does not have observable change until the start of disturbances for the experimental run.
Tapping begins about thirty minutes after the pile was formed.

\begin{figure}[h]
\centering
\includegraphics[scale=0.7]{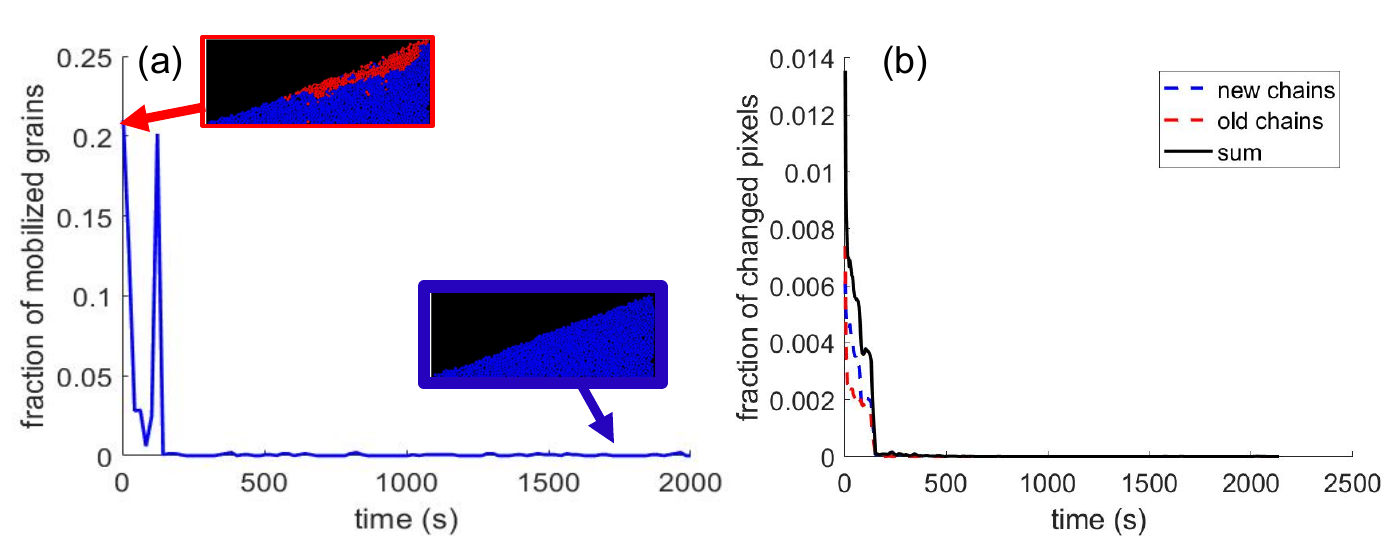}
\caption{\textbf{Pile relaxation:} (a) Fraction of grains that move above a displacement threshold of an average grain radius plotted as a function of time. The images illustrate grains that move above the threshold at the noted times. (b) Corresponding changes in the force network as a function of time.
}
\label{fig:pile_formation}
\end{figure}

\vspace{1cm}
\section{Sample Movie of an experimental run}
A typical movie of the behavior described in the main text is included with the supplemental material.
Images were taken every three seconds and played back at ten frames per second.
The spatial resolution is 8 pixels per millimeter.
In this example, one observes small rearrangements, creep among surface grains, and an avalanche event.

\vspace{1cm}
\section{Disturbance profile}
\begin{figure}[h]
\centering
\includegraphics[width = 0.5\linewidth]{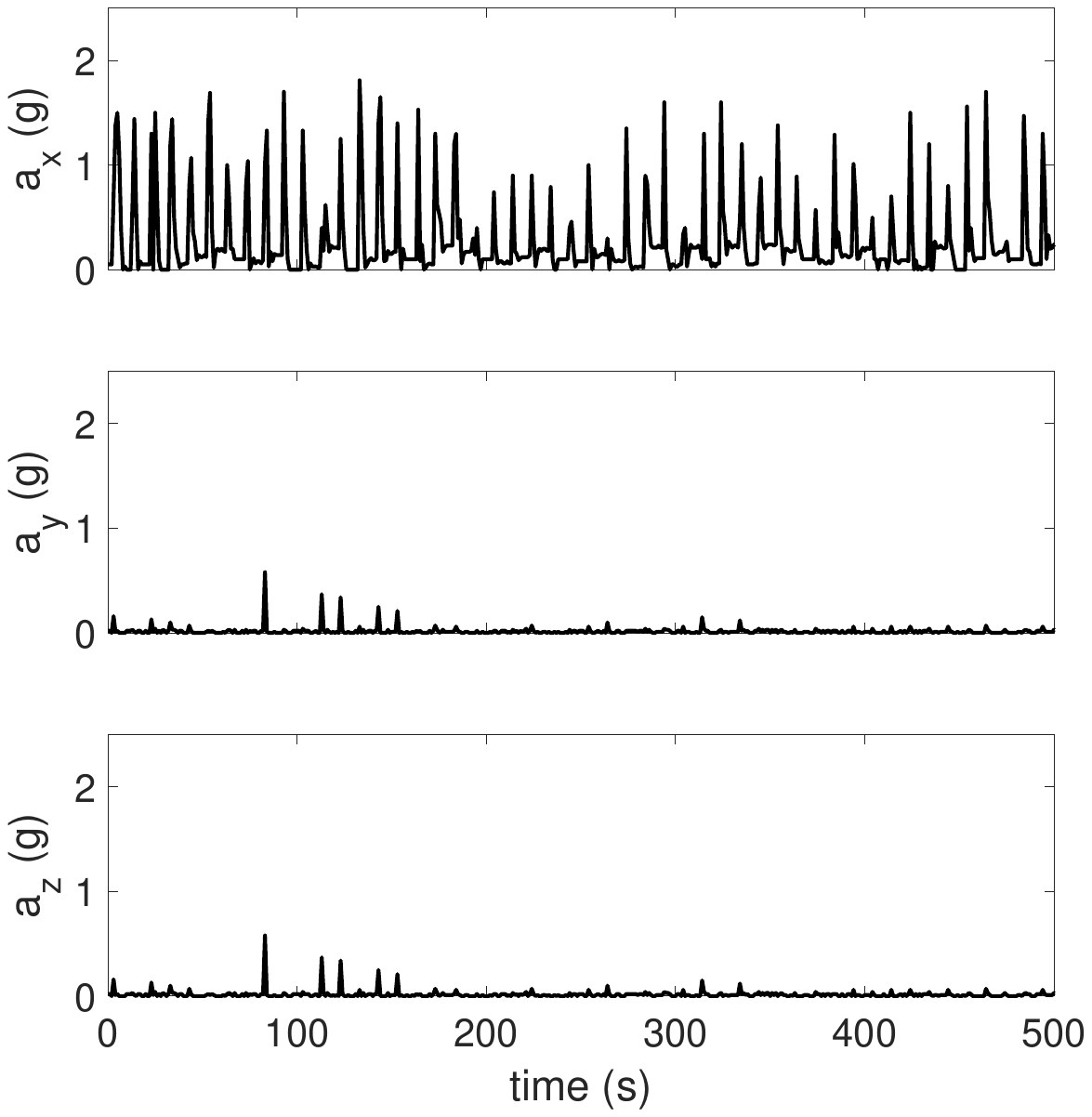}
\caption{\textbf{Tapping measurements:} Acceleration measurements (in units of g) as a function of time for taps from the mechanical arm on the pile apparatus. Measurements were performed using an accelerometer ADXL 337 attached to the arm and was read by an Arduino.}
\label{fig:tapping}
\end{figure}

% describe how a tap works
Here we provide additional details about the tapping protocol. 
Supplemental figure \ref{fig:tapping} shows the acceleration experienced by the pile due to the periodic taps by a mechanical arm.
All data presented in the main text are for 10-second tap intervals.
The tapping rate does not seem to increase the likelihood of an avalanche.
The initial surface angle of the pile is a better predictor of an ensuing avalanche.

\newpage
\section{Displacement and force measurements for additional experimental runs}
In the main text, we report on the dynamical and structural signatures that occur as a granular pile creeps and avalanches.
In particular, there is a stress precursor that indicates a large grain motion event.
These qualitative dynamics are similar for other runs that involve an avalanche; another example is shown in the supplemental figure \ref{fig:extraruns}(a).
If more than 10 percent of the grains of the pile move above the distance threshold within a five-frame increment, 
we label the event as an avalanche. 
This definition is then supported by dynamic and geometric signatures.
Note that the experimental run involving creep only (Fig. \ref{fig:extraruns}(b)) does not exhibit a large decrease of stress profile that coincides with significant grain motion.

\begin{figure}[h]
\centering
\includegraphics[width = \linewidth]{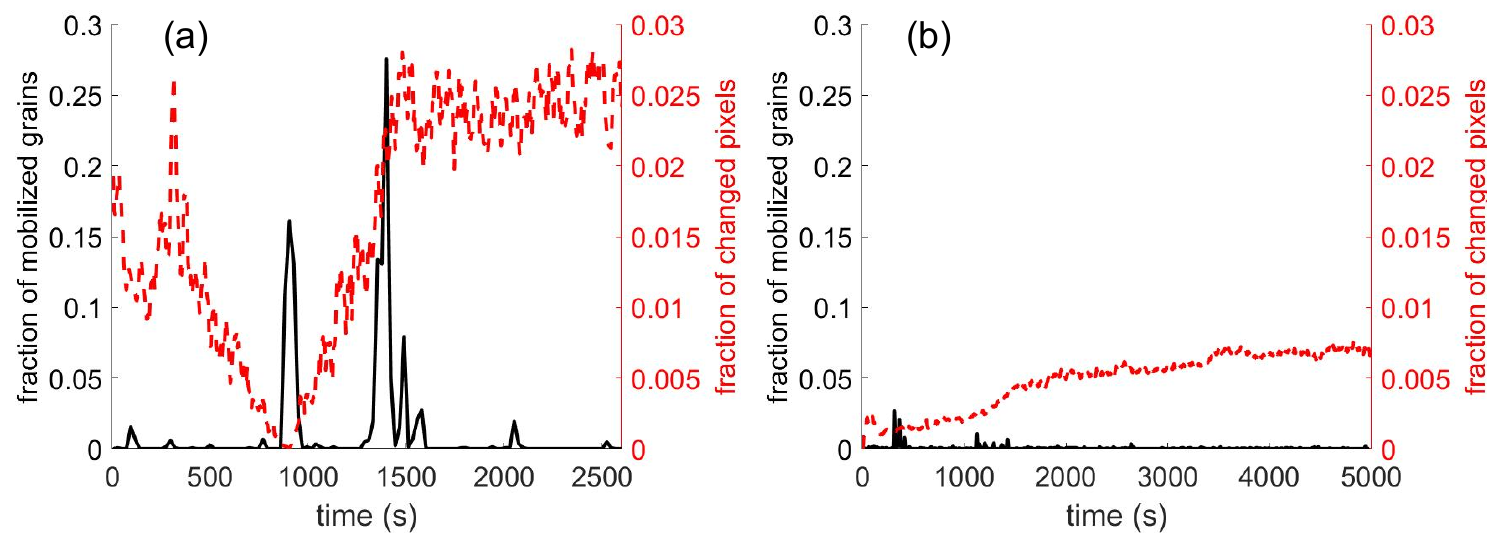}
\caption{\textbf{Additional runs:} Fraction of black and white pixels as determined from difference of polarized light images plotted with number of grains that move above a distance threshold of one particle radius, showing how stress redistribution leads to (a) failure and (b) creep.}
\label{fig:extraruns}
\end{figure}

\vspace{1cm}
\section{Void analysis}

We observe that grain displacement events coincide with shifts in the unoccupied volume. 
We provide an additional run in supplemental figure 4(a).
To further test this observation, we analyze the void space as a function of time in a pile formation process where large grain rearrangements occur at the beginning (due to the deposition of grains) and negligible grain motion is observed at later times (absence of external disturbances). 
This test shows: (i) a clear decrease in unoccupied volume due to the deposition of grains and (ii) unchanging unoccupied volume when there is negligible grain movement at later times (see supplemental Fig. 4(b)).

\begin{figure}[h]
\centering
\includegraphics[width =0.5\linewidth]{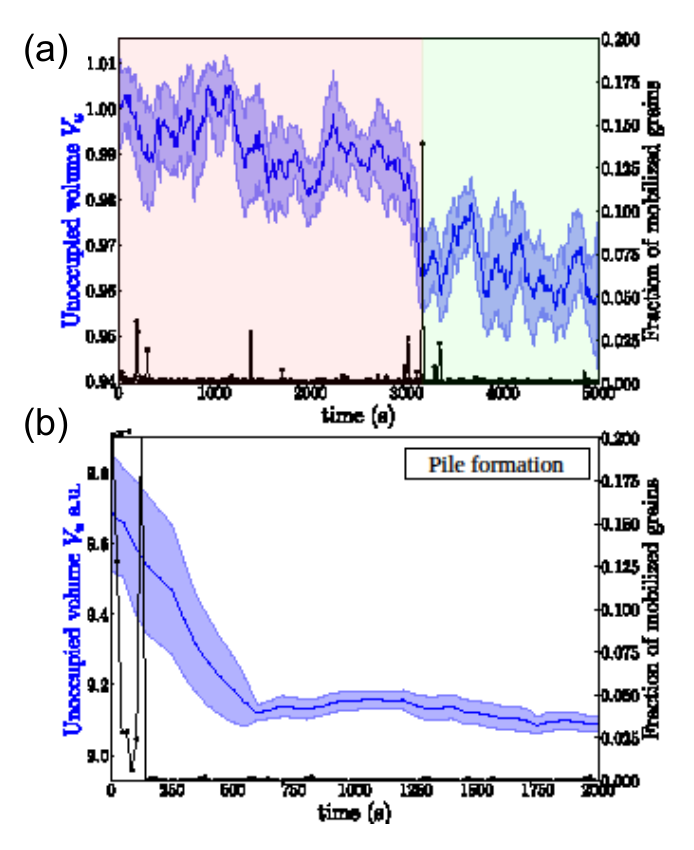}
\caption{(a) Additional void data as a function of time for another avalanche run. (b) Reference data using a grain time-series data from pile formation (see also supplemental figure \ref{fig:pile_formation}). When there is no movement of grains the unoccupied volume remains constant. }
\label{fig:Suppvoids}
\end{figure}

\bibliography{references.bib}